\documentclass[useAMS]{mn2e}
\usepackage{threeparttable}
\usepackage{multicol}
\usepackage{aas_macros,graphicx,times,multirow,amsmath,amssymb,color,longtable}
\usepackage[T1]{fontenc}

\title[The magnetic CV: RX\,J2015.6$+$3711]{Multiwavelength study of RX\,J2015.6$+$3711: a magnetic cataclysmic variable with a 2-hr spin period}

\author[F.~Coti Zelati et al.]{F.~Coti Zelati,$^{1,2,3}$\thanks{E-mail: francesco.cotizelati@brera.inaf.it} N.~Rea,$^{2,4}$ S.~Campana,$^{3}$ 
D. de~Martino,$^{5}$ A.~Papitto,$^{4}$ \newauthor S.~Safi-Harb$^{6}$ and D.~F.~Torres$^{4,7}$
\smallskip\\
$^{1}$ Universit\`a dell' Insubria, via Valleggio 11, I-22100 Como, Italy\\
$^{2}$ Anton Pannekoek Institute for Astronomy, University of Amsterdam, Postbus 94249,  NL-1090-GE Amsterdam, The Netherlands\\
$^{3}$ INAF -- Osservatorio Astronomico di Brera, via Bianchi 46, I-23807 Merate (LC), Italy\\
$^{4}$ Instituto de Ciencias de l'Espacio (ICE, CSIC--IEEC), Carrer de Can Magrans, S/N, E-08193 Barcelona, Spain\\
$^{5}$ INAF -- Osservatorio Astronomico di Capodimonte, Salita Moiariello 16, I-80131 Napoli, Italy\\
$^{6}$ Department of Physics and Astronomy, University of Manitoba, Winnipeg, MB R3T 2N2, Canada\\
$^{7}$ Instituci\'o Catalana de Recerca i Estudis Avan\c{c}ats (ICREA), E-08010 Barcelona, Spain\\
}

\date{\today}
\pagerange{\pageref{firstpage}--\pageref{lastpage}} 
\pubyear{2015}

\def\ltsima{$\; \buildrel < \over \sim \;$}
\def\lsim{\lower.5ex\hbox{\ltsima}}
\def\gtsima{$\; \buildrel $\geq$ \over \sim \;$}
\def\gsim{\lower.5ex\hbox{\gtsima}}

\newcommand{\be}{\begin{equation}}
\newcommand{\en}{\end{equation}}

\def\nh{\hbox{$N_{\rm H}$}}
\def\flux {\mbox{erg cm$^{-2}$ s$^{-1}$}}
\def\lum {\mbox{erg s$^{-1}$}}
\def\cv{\mbox{RX\,J2015.6$+$3711}}

\def\j1023{\mbox{PSR\,J1023$+$0038}}
\def\xss{\mbox{XSS\,J12270$-$4859}}

\def\LaTeX{L\kern-.36em\raise.3ex\hbox{a}\kern-.15em
    T\kern-.1667em\lower.7ex\hbox{E}\kern-.125emX}

\newcommand{\axaf}{{\em Chandra}}
\newcommand{\ros}{{\em ROSAT}}

\newcommand{\xmm}{{\em XMM--Newton}}

\newcommand{\swift}{{\em Swift}}
\newcommand{\fermi}{{\em Fermi}}

\begin{document}

\label{firstpage}
\pagerange{\pageref{firstpage}--\pageref{lastpage}}
\maketitle

\begin{abstract}
The X-ray source \cv\ was discovered by \ros\ in 1996 and recently proposed to be a cataclysmic 
variable (CV). Here we report on an \xmm\, observation of \cv\ performed in 2014, where we detected 
a coherent X-ray modulation at a period of $7196 \pm 11$~s and discovered other significant ($>6\sigma$) 
small-amplitude periodicities which we interpret as the CV spin period and the sidebands of a possible 
$\sim$12\,hr periodicity, respectively. The 0.3--10~keV spectrum can be described by a power law 
($\Gamma = 1.15\pm0.04$) with a complex absorption pattern, a broad emission feature at $6.60\pm0.01$~keV, 
and an unabsorbed flux of $(3.16\pm0.05) \times 10^{-12}$ \flux. We observed a significant spectral 
variability along the spin phase, which can be ascribed mainly to changes in the density of a partial 
absorber and the power law normalization. Archival X-ray observations carried out by the \axaf\ satellite, 
and two simultaneous X-ray and UV/optical pointings with \swift, revealed a gradual fading of the source 
in the soft X-rays over the last 13 years, and a rather stable X-ray--to--optical flux ratio 
($F_{\mathrm{X}}/F_{\mathrm{V}} \approx 1.4-1.7$). Based on all these properties, we identify this source 
with a magnetic CV, most probably of the intermediate polar type. The 2\,hr spin period makes \cv\, 
the second slowest rotator of the class, after RX\,J0524$+$4244 ("Paloma"; $P_{\rm spin}\sim2.3$~hr). Although 
we cannot unambiguously establish the true orbital period with these observations, \cv\ appears to be a key 
system in the evolution of magnetic CVs.
\end{abstract}

\section{Introduction}

Cataclysmic variables (hereafter CVs) are interacting binary systems in which a white dwarf 
(WD) accretes matter from a late-type low mass main sequence star through Roche lobe 
overflow. Typical orbital periods of these systems are of a few hours (see Warner 1995 for a 
review). About 20--25\% of the known CVs harbour WDs with magnetic field in the 
$10^5-10^8$\,G range, and are called magnetic CVs (mCVs; Ferrario, de Martino 
\& G\"{a}nsicke 2015). The mCVs are further classified as polars and intermediate polars 
(IPs) based on the strength of their magnetic field. The former systems are 
characterized by very high magnetic fields ($B \sim 10^7 - 10^{8}$~G) that 
are able to synchronise the WD spin with the orbital period to a very high degree 
($P_{{\rm spin}} \simeq P_{\rm{orb}}$). The latter are believed to possess weaker fields 
($B\lesssim10^{7}$~G) because of the asynchronous rotation of the WD. They have in 
fact spin periods of a few hundreds of seconds and orbital periods of a few hours, with 
a spin--to--orbit period ratio P$_{{\rm spin}}$/P$_{{\rm orb}}$ $\approx$ 0.05 - 0.15 
(Norton, Wynn \& Somerscales 2004).

\cv\ was discovered by the High Resolution Imager on board  \ros\ in 1996 August, during a 
survey of the gamma-ray source 3EG\,J2016$+$3657 in the EGRET catalog (Halpern et al. 2001). 
It has optical and near-infrared magnitudes of $R\sim17.5$, $J = 15.54\pm0.08$, $H = 15.04\pm0.11$ 
and $K = 14.79\pm0.15$ (see Halpern et al. 2001 and the Two-Micron All-Sky Survey 
catalog\footnote{http://www.ipac.caltech.edu/2mass/releases/second/}). 
\cv\ lies in the error box of the variable \fermi\ source 3FGL\,J2015.6$+$3709 (Acero et al. 2015) 
and within a crowded region of high-energy emitting sources. The blazar B2013$+$370 is only 
$\sim1.6$ arcmin away and it is also compatible with the position of the \fermi\ source (Bassani et 
al. 2014). Furthermore, the supernova remnant CTB~87 is located at an angular separation of 
about 5.2~arcmin from the source (Matheson, Safi-Harb \& Kothes 2013).

\begin{table*}
\begin{minipage}{16.cm}
\centering \caption{Journal of the X-ray observations used in this work.} 
\label{tab:obslog}
\begin{tabular}{@{}lcclcc}
\hline
Satellite 		& Instrument 				& Obs.\,ID  			& Date 				& Exposure$^{a}$ 		& Mode$^{b}$\\
			&  						& 					&					& (ks) 				& \\
\hline
\axaf 		& ACIS-S					& 1037				& 2001 Jul 08 			& 17.8 				& TE FAINT (3.241\,s)\\
\swift			& XRT 					& 00035639003		& 2006 Nov 17			& 7.4 				& PC (2.507\,s)\\
\axaf 		& ACIS-I		 			& 11092				& 2010 Jan~16--17 		& 69.3 				& TE VFAINT (3.241\,s)\\
\swift			& XRT 					& 00041471002		& 2010 Aug~06--10		& 7.3 				& PC (2.507\,s)\\
\xmm		& pn / MOS\,1 / MOS\,2 		& 0744640101 			& 2014 Dec~14--16		& 108.2 / 122.3 / 122.3	& FF (73.4\,ms) / FF (2.6\,s) / FF (2.6\,s) \\
\hline
\end{tabular}
\begin{list}{}{}
\item[$^{a}$] Deadtime corrected on-source time.
\item[$^{b}$] TE: Timed Exposure, FAINT: Faint telemetry format, VFAINT: Very Faint telemetry format, FF: Full Frame, LW: Large Window, PC: Photon Counting;  the temporal resolution is given in parentheses. 
\end{list}
\end{minipage}
\end{table*}

The spatial coincidence of \cv\ with the GeV source has opened the new possibility (Bassani 
et al. 2014) that it may belong to the class of the recently discovered transitional 
millisecond pulsars (TMPs). TMPs are neutron star low-mass X-ray binary systems (LMXBs) 
that are observed to switch between accretion and rotation-powered emission on timescales 
ranging from a few weeks to a few years (see e.g. Papitto et al. 2013). The possibility of \cv\ 
being such a system is not remote if one considers the cases of the sources \j1023\ and \xss. 
These were initially classified as mCVs (Thorstensen \& Armstrong 2005; Masetti et al. 2006) 
and subsequently recognized to be gamma-ray emitters with a 0.1--10~GeV luminosity of a few 
$10^{33}$~\lum, comparable to that in the X-rays (Stappers et al. 2014; de Martino et al. 2010, 
2013). Multiwavelength observations over the last years led observers to identify these 
sources as TMPs (see Bogdanov et al. 2015 and references therein; de Martino et al. 2015 and 
references therein).

However, while this work was in preparation, Halpern \& Thorstensen (2015) suggested that the 
source might be a mCV of the polar type, on the basis of the characteristics of its optical spectrum 
and the discovery of an energy dependent 2-hr X-ray modulation in archival \axaf\, data, on which 
we report independently here in detail using also a long \xmm\, observation.

Here we attempt to better assess the nature of \cv\ through a detailed analysis of a recent \xmm\ 
observation, and a reanalysis of archival \axaf\ and \swift\ observations. We describe the X-ray and 
UV/optical observations and present the results of our data analysis in Section~\ref{obs} and \ref{uv}. 
We discuss our results in Section~\ref{discussion}. Conclusions follow in Section~\ref{conclusion}.

\section{X-ray observations and data analysis}
\label{obs}

The field of \cv\ was observed multiple times by the X-ray imaging instruments on 
board the \xmm, \axaf\ and \swift\ satellites. A summary of the observations used in 
our study is reported in Table~\ref{tab:obslog}.

\subsection{\xmm}

A deep \xmm\ observation (PI: Safi-Harb) was carried 
out using the European Photon Imaging Cameras (EPIC),
starting on 2014 December 14 (see Table~\ref{tab:obslog})
and with CTB~87 placed at the aim point. The pn (Str\"{u}der 
et al. 2001) and the two MOS (Turner et al. 2001) CCD cameras 
were configured in full-frame window mode. The medium optical 
blocking filter was positioned in front of the cameras.

We processed the raw data files using the \textsc{epproc} (for 
pn data) and \textsc{emproc} (for MOS data) tasks of the \xmm\ 
Science Analysis System (\textsc{sas}\footnote{http://xmm.esac.esa.int/sas/}, 
version 14.0), with the most up to date calibration files available.
The data were affected by strong soft-proton flares of solar origin.
For the timing analysis we decided to dynamically subtract the scaled 
background in each bin of the source light curves binned at 10~s. 
For the spectral analysis we built the light curve of the entire field 
of view and discarded episodes of flaring background using intensity 
filters. This reduced the effective exposure time to approximately 77.8, 
107.3 and 106.8 ks for the pn, MOS\,1 and MOS\,2, respectively.

\begin{figure}
\begin{center}
\includegraphics[width=0.48\textwidth]{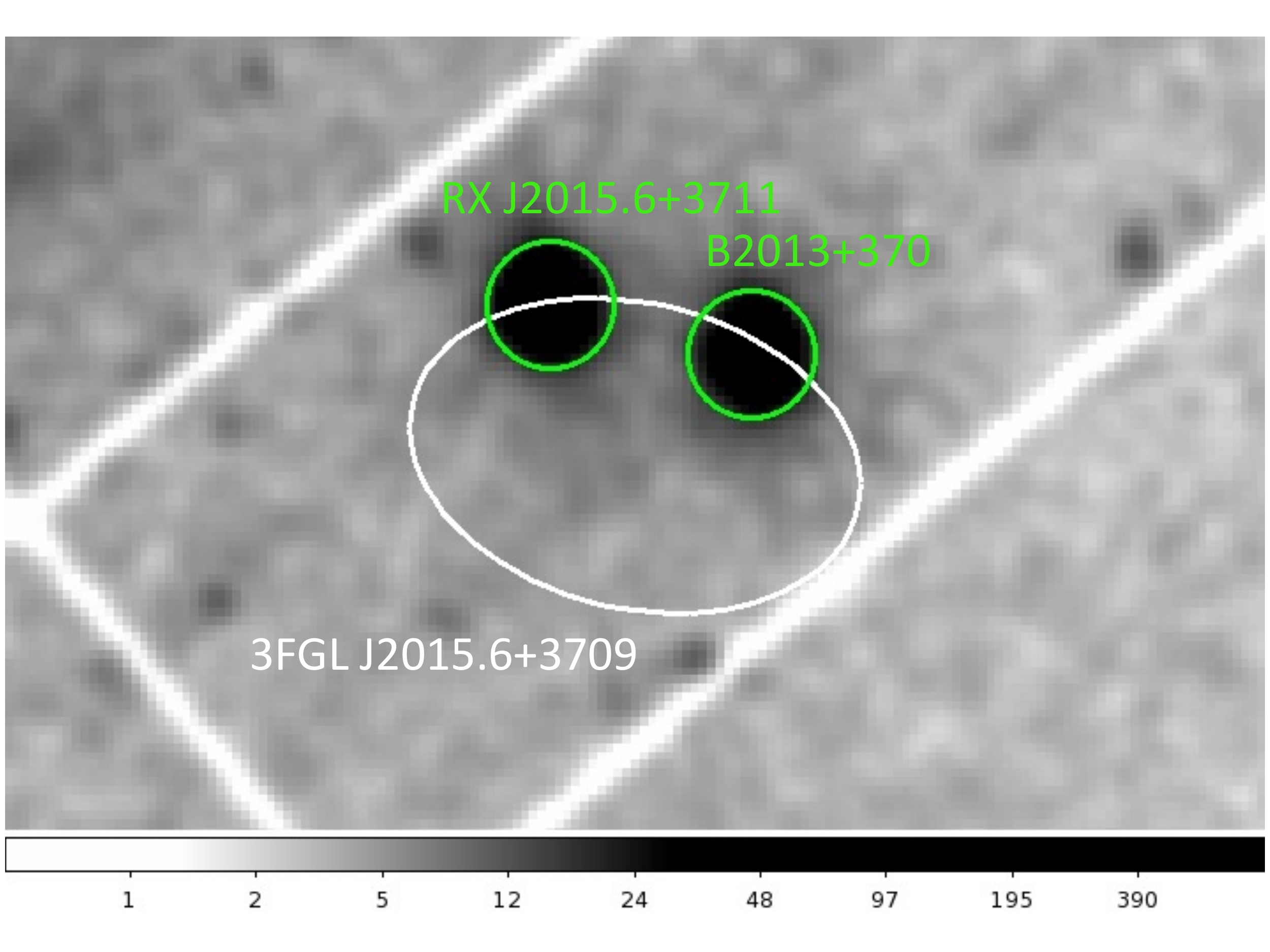}
\vspace{-0.5cm}
\caption{\xmm\ EPIC-pn 0.3--10~keV image of the field around \cv. The field 
size is about $12 \times 7.5$~arcmin$^2$. North is up, east to the left. Both 
\cv\ and the blazar B2013$+$370 were located on CCD 5 and their positions are 
marked with green circles, each with a radius of 30 arcsec. The white ellipse is 
centered at $\rm RA=20^h15^m33\fs6$, $\rm Dec= +37^\circ10'12\farcs2$ 
(J2000.0) and represents the position (at the 95\% confidence level) of the source 
3FGL\,J2015.6$+$3709 in the \fermi\ Large Area Telescope 4-year point source 
catalog (Acero et al. 2015; see also Kara et al. 2012). (See the online version of 
the article for the colour figure.)
}
\label{fig:pn_fov}
\end{center}
\end{figure}

The field of \cv\ observed by the pn camera is shown in Fig.~\ref{fig:pn_fov}.
The image was created by selecting events with \textsc{pattern}=0, 
(\textsc{flag} \& 0x2fa002c)=0 for the 0.3--0.5~keV range, \textsc{pattern}
$\leq$4, (\textsc{flag} \& 0x2fa002c)=0 for the 0.5--1.0~keV range 
and \textsc{pattern}$\leq$4, (\textsc{flag} \& 0x2fa0024)=0 for the 
1.0--10~keV range, to remove any traces of hot pixels and bad 
columns, and was also smoothed with a Gaussian filter with a kernel 
radius of 3 pixels (one EPIC-pn pixel corresponds to about 4.1 arcsec).

We extracted the source photons from a circular region centered 
at the optical position of the source ($\rm RA=20^h15^m36\fs98$, 
$\rm Dec= +37^\circ11'23\farcs2$ (J2000.0); see Halpern et al. 
2001) and with a radius of 30 arcsec, to avoid contamination from the 
closeby blazar. The background was extracted from a similar region, 
far from the source location and on the same CCD (see Bassani et 
al. 2014 for an overview of the nearby X-ray emitting sources). We 
converted the photon arrival times to Solar System barycentre reference 
frame using the \textsc{sas} task \textsc{barycen} and restricted our 
analysis to photons having energies between 0.3 and 10 keV. We 
verified that data were not affected by pile-up through the \textsc{epatplot} 
script. The 0.3--10~keV average source net count rates are $(2.74 \pm 0.02) 
\times 10^{-1}$, $(8.98 \pm 0.09) \times 10^{-2}$ and $(8.83 \pm 0.09) 
\times 10^{-2}$ counts~s$^{-1}$ for the pn, MOS\,1 and MOS\,2, respectively.

We applied the standard filtering procedure in the extraction of the spectra, 
retaining only events optimally calibrated for spectral analysis (\textsc{flag} = 0) 
and with \textsc{pattern} $\leq4(12)$ for the pn (MOS) data. We generated the 
corresponding redistribution matrices and ancillary response files with the 
\textsc{rmfgen} and \textsc{arfgen} tools, respectively, and grouped the 
background-subtracted spectra to have at least 100 counts in each spectral bin.

\begin{figure*}
\begin{center}
\includegraphics[width=1\textwidth]{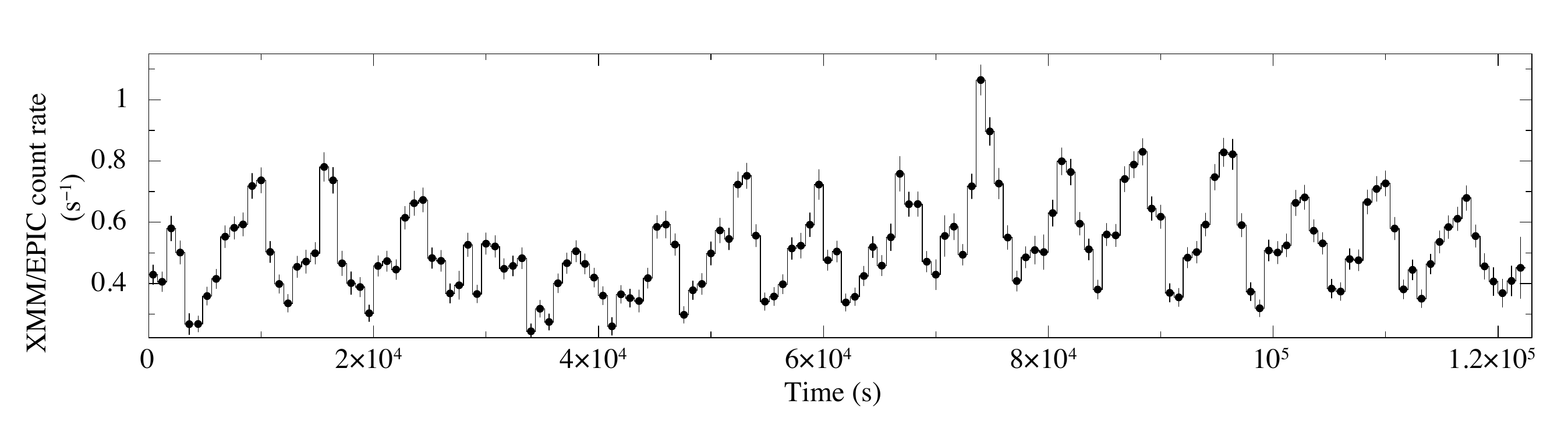}
\vspace{-0.5cm}
\caption{0.3--10~keV background-subtracted and exposure-corrected light curve 
of \cv\ obtained with the \xmm\ EPIC cameras with a binning time of 800~s.}
\label{fig:lcurve}
\end{center}
\end{figure*}

\subsubsection{Timing analysis} 
\label{timing}

\begin{figure*}
\begin{center}
\includegraphics[width=0.5\textwidth]{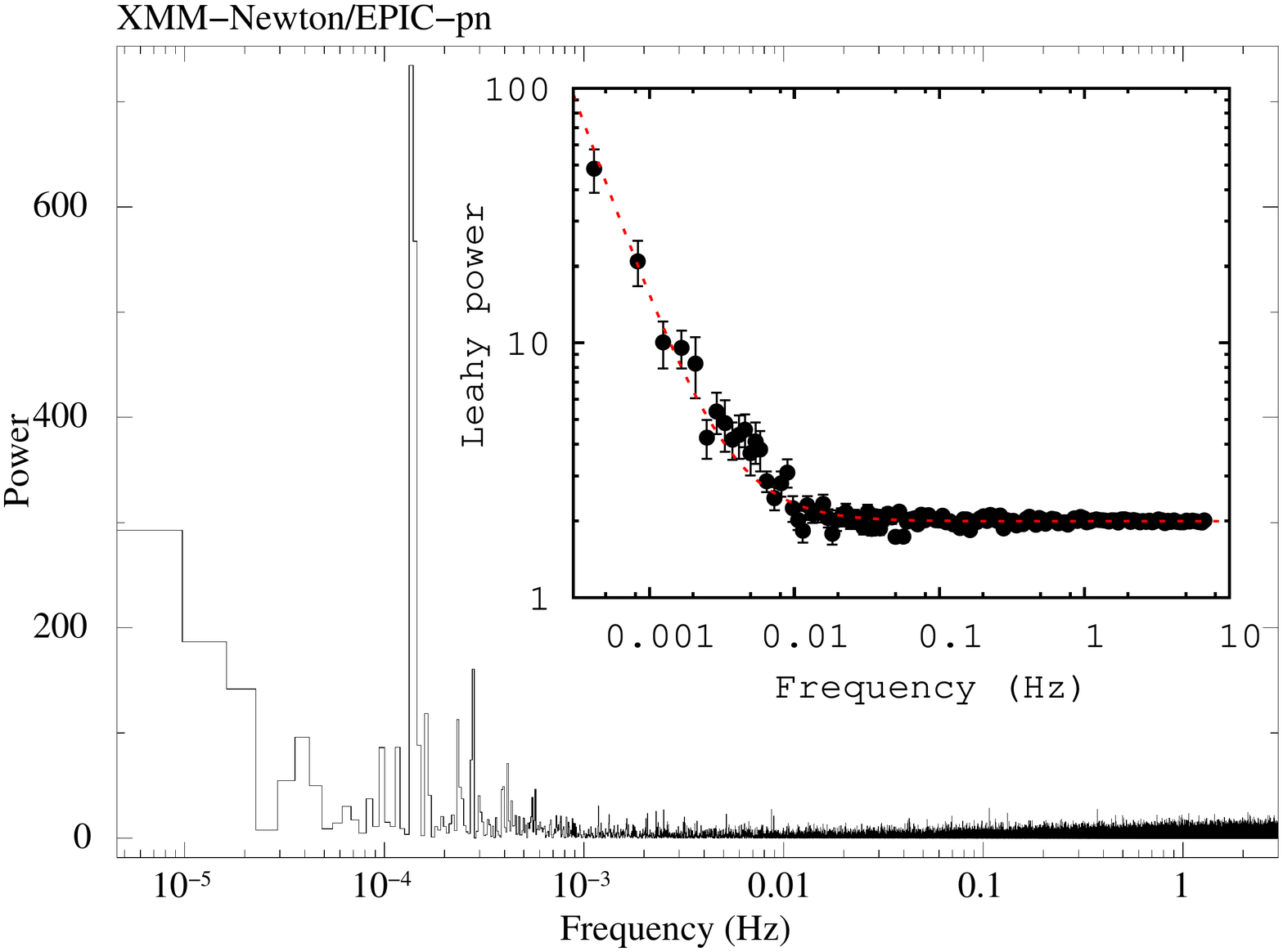}
\includegraphics[width=0.495\textwidth]{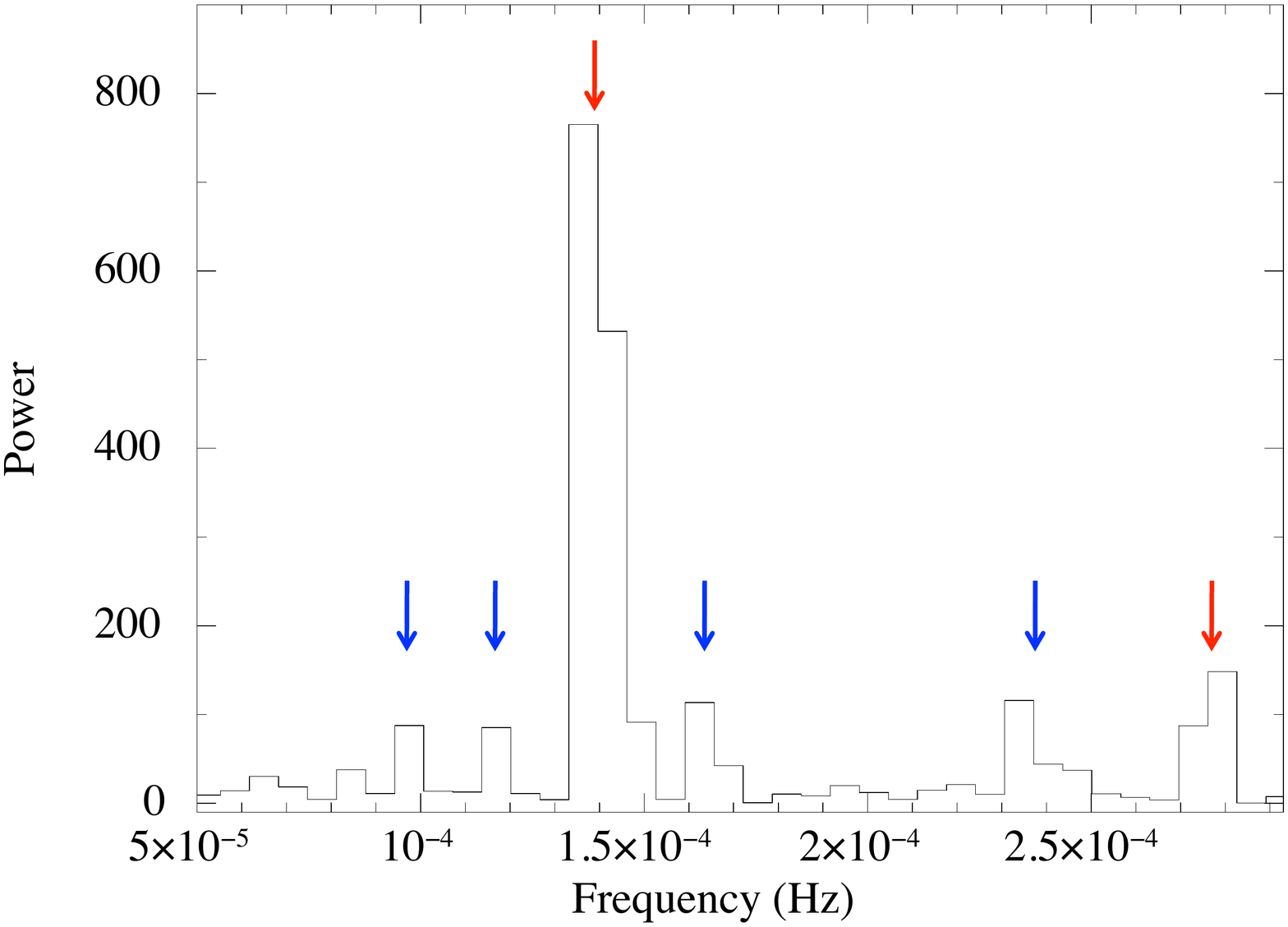}
\vspace{-0.5cm}
\caption{Left-hand panel: 0.3--10~keV power spectrum of \cv\ obtained from \xmm\ 
EPIC-pn data. Inset: Fourier power spectral density of \cv. The red dashed line represents 
the modeling of the white and red noise components, i.e. the sum of a constant and a 
power law function: $P(\nu)=K+C\nu^{-\beta}$, with $K=1.99(1)$~Hz$^{-1}$ and 
$\beta=1.59(8)$. Right-hand panel: 0.3--10~keV power spectrum of \cv\ restricted to the 
frequency range between $5 \times 10^{-5}$ and $3 \times 10^{-4}$~Hz. The main peak 
and that of the second harmonic are highlighted with red arrows, whereas the other significant 
power peaks are marked with blue arrows. All these peaks are detected at a significance 
level above $6\sigma$ (see the text). (See the online version of the article for the colour figure.)
}
\label{fig:powspec}
\end{center}
\end{figure*}

The 0.3--10~keV background-subtracted and exposure-corrected light curve of \cv\
is reported in Fig.~\ref{fig:lcurve}. It was generated by combining the time series from the 
EPIC cameras during the periods when all three telescopes acquired data 
simultaneously, using the \textsc{epiclccorr} tool of \textsc{sas} and other tasks in the 
\textsc{ftools} package (Blackburn 1995). The light curve clearly shows a periodic 
modulation. We computed a Fourier transform of the pn light curve and found indeed 
a prominent peak at a frequency of $\sim1.39 \times 10^{-4}$~Hz (at a significance 
level of about 27$\sigma$, estimated taking into account the presence of the underlying 
white noise component and the number of independent Fourier frequencies examined 
in the power spectrum; see the left-hand panel of Fig.~\ref{fig:powspec}). A search for 
periodicities at millisecond periods was precluded owing to the temporal resolution of 
the pn camera in full frame readout mode, which implies a Nyquist limiting frequency 
of about 6.8~Hz. We show in the inset of Fig.~\ref{fig:powspec} the Fourier power spectral 
density of \cv, produced by calculating the power spectrum into 2405~s-long consecutive 
time intervals, averaging the 48 spectra so evaluated and rebinning geometrically 
the resulting spectrum with a factor 1.05. A closer inspection of the power spectrum 
around the frequency of the main peak revealed excess of power up to the second 
harmonic and also at frequencies of about $9.4 \times 10^{-5}$, $1.20 \times 10^{-4}$, 
$1.65  \times 10^{-4}$ and $2.55 \times 10^{-4}$~Hz (in all cases at a significance 
level larger than $6\sigma$; see the right-hand panel of Fig.~\ref{fig:powspec}). 
The power spectra filtered in different energy intervals show no evidence for significant 
peaks above $\sim 3$ keV.

To refine our period estimate for the main peak in the power spectrum, we 
performed an epoch-folding search of the light curve by binning the profile into 16 
phase bins and searching with a period resolution of 13 s. We then fitted the peak 
in the $\chi^2$ versus trial period distribution as described by Leahy (1987) and 
derived a best period $P=7196 \pm 11$~s (at the 90\% confidence level), 
consistent within the errors with that reported (independently) by Halpern \& Thorstensen 
(2015).

\begin{figure*}
\begin{center}
\vspace{-0.8cm}
\includegraphics[width=1.05\textwidth]{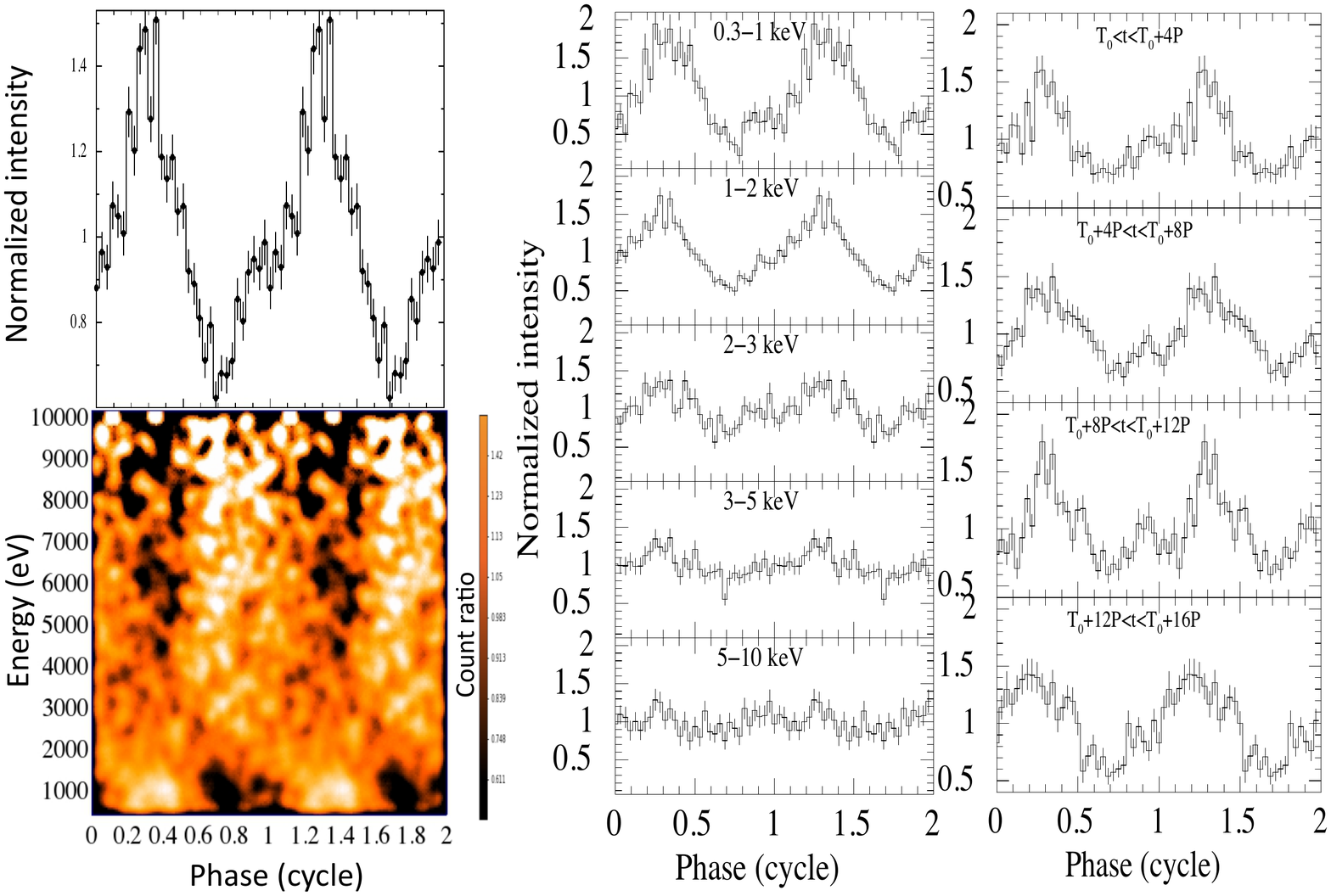}
\vspace{-1.5cm}
\caption{Left-hand panel, top: 0.3--10~keV background-subtracted and 
exposure-corrected light curve of \cv\ (from \xmm/EPIC-pn data) folded 
on the period $P$=7196~s and sampled in 32 phase bins. Epoch 
T$_0$=57\,005.6889669 MJD was used as reference. Left-hand panel, 
bottom: normalized energy versus phase image for the pn data of \cv\ 
(see the text for details). Middle panel: 32-bin pulse profiles in five different 
energy bands. Energy increases from top to bottom. Right-hand panel: 
32-bin pulse profiles in four different time intervals (0.3--10~keV). Time 
increases from top to bottom. Two cycles are shown in all cases for better 
visualization. (See the online version of the article for the colour figure.)
}
\label{fig:profiles}
\end{center}
\end{figure*}

The pn light curve folded on $P$ is shown in the left-hand panel of Fig.~\ref{fig:profiles}. 
The profile is asymmetric and is characterized by a prominent peak 
in the phase interval $\sim 0.1-0.4$, followed by a minimum in the $0.6-0.9$ range. 
We also produced an energy versus phase image by binning the source counts into 
100 phase bins and 100-eV wide energy channels and normalizing to the phase-averaged 
energy spectrum and pulse profile. A lack of counts is evident at energy $\lesssim 2$~keV 
and in the phase interval corresponding to the minimum of the profile (see again 
the left-hand panel of Fig.~\ref{fig:profiles}). 

The profiles are approximately phase-aligned at different energies within statistical 
uncertainties, but their morphology changes as a function of energy (see the middle 
panel of Fig.~\ref{fig:profiles}). In particular we observe a secondary hump in the phase 
interval 0.8--1.1 in the softest band (0.3--1~keV), a feature that is nearly absent in the 
1--2~keV interval. To assess the significance of the observed pulse shape variations 
as a function of energy, we compared the 0.3--1~keV and 1--2~keV folded profiles 
using a two-sided Kolmogorov--Smirnov test (Peacock 1983; Fasano \& Franceschini 
1987). The result shows that the difference between these energy ranges is highly 
significant: the probability that the two profiles do not come from the same underlying 
distribution is in fact $\sim4\times10^{-7}$, corresponding to a significance of $\sim 5.3 
\sigma$.

We also analyzed the temporal evolution of the pulse shape in the 0.3--10~keV 
energy band by dividing the entire duration of the pn observation into four consecutive 
time intervals, each of length 4$P$, and folding the corresponding light curves on 
$P$. As shown in the right-hand panel of Fig.~\ref{fig:profiles}, the shape of the profile 
clearly changes as time elapses.

We evaluated the pulsed fractions by fitting a constant plus two sinusoidal functions to 
the 32-bin folded light curves and considering the semi-amplitude measured from the 
fundamental frequency component (the sinusoidal periods were fixed to those of the 
fundamental and second harmonic components). The inclusion of higher harmonic 
components in the fits was not statistically needed, as determined by means of an 
$F$--test. The 0.3--10~keV pulsed fractions are $29 \pm 1$, $29 \pm 2$ and $30 \pm 2$\% 
for the pn, MOS\,1 and MOS\,2 data sets, respectively (uncertainties are reported 
at the 90\% confidence level). The amplitude of the modulation decreases as the 
energy increases: for pn data we calculated pulsed fractions of $62 \pm 4$, $44 \pm 2$, 
$22 \pm 3$, $17 \pm 3$ and $9 \pm 4$\% in the 0.3--1, 1--2, 2--3, 3--5 and 5--10~keV 
energy ranges, respectively. The pulsed fraction is however consistent with the average 
value over the entire duration of the observation. 

\begin{figure}
\includegraphics[width=0.45\textwidth]{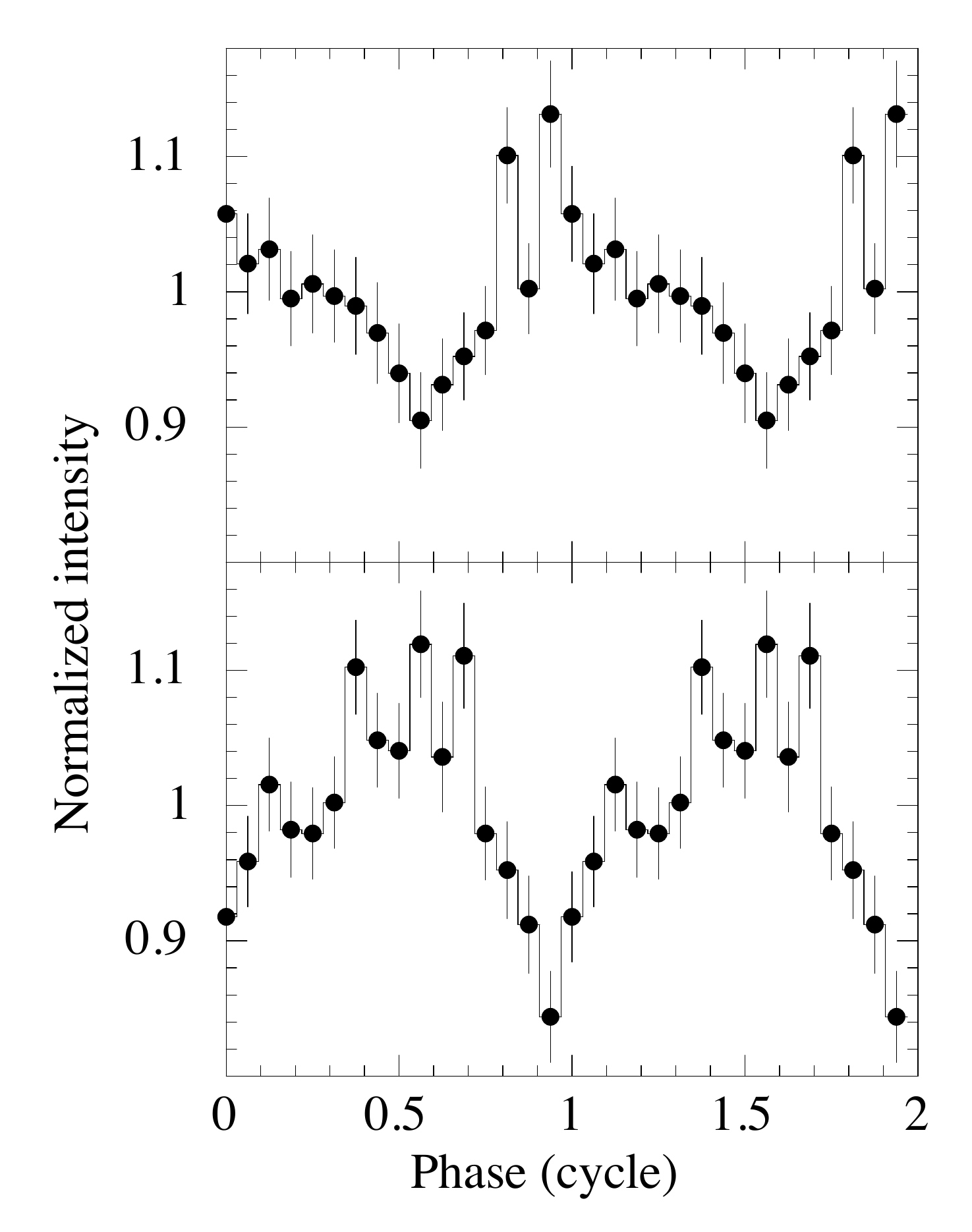}
\caption{0.3--10~keV background-subtracted and 
exposure-corrected light curve of \cv\ folded 
on the periods $P$=8332~s (upper panel) and 
$P$=6054~s (lower panel), and sampled in 16 phase bins.
Two cycles are shown for better visualization.}
\label{fig:sideband}
\end{figure}

In Fig.~\ref{fig:sideband} we show the 0.3--10~keV profiles folded at the best periods 
corresponding to the frequencies $1.20 \times 10^{-4}$ and $1.65  \times 10^{-4}$~Hz (i.e. 
$8332 \pm 14$~s and $6052 \pm 6$~s, respectively). The modulation amplitudes are 
$6 \pm 1$ and $8 \pm 1$\%, respectively.

\subsubsection{Phase-averaged spectral analysis}
\label{spectrum}

We fitted the spectra of the three EPIC cameras together in the 
0.3--10~keV energy range using the \textsc{xspec}\footnote{http://heasarc.gsfc.nasa.gov/xanadu/xspec/} 
spectral fitting package (v. 12.9.0; Arnaud 1996). We adopted a 
set of different single-component models: a blackbody 
(\textsc{bbodyrad} in \textsc{xspec} notation), an optically-thin 
thermal bremsstrahlung, an accretion disk consisting of multiple 
blackbody components (\textsc{diskpbb}) and a power law 
(\textsc{pegpwrlw}). To describe the absorption by the interstellar 
medium, we used the Tuebingen-Boulder model (\textsc{tbabs}), 
with photoionization cross-sections from Verner et al. (1996) and 
solar chemical abundances from Wilms, Allen \& McCray (2000). 
We included an overall normalization factor to account for calibration 
uncertainties among the three different X-ray detectors\footnote{This 
factor was fixed to 1 for the pn spectrum, and left free to vary for both 
the MOS\,1 and MOS\,2 spectra.} and tied up all the parameters across 
the three data sets. In the following we will quote all the uncertainties 
at a 90\% confidence level for a single parameter of interest 
($\Delta \chi^2 = 2.706$), unless otherwise specified.

\begin{figure}
\begin{center}
\includegraphics[width=0.5\textwidth]{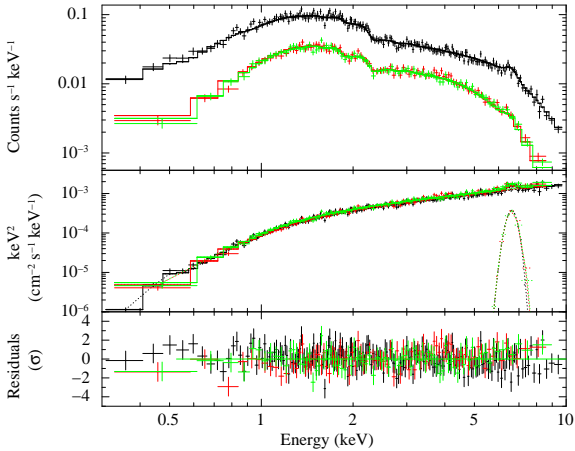}
\vspace{-0.5cm}
\caption{0.3--10~keV average spectrum of \cv\ extracted from 
\xmm\ EPIC data. The solid line represents the best-fitting model, i. e.
the superposition of an absorbed power law and a Gaussian feature in 
emission. $E^2\times f(E)$ unfolded spectra (middle panel) and 
post-fit residuals in units of standard deviations (bottom panel) are 
also shown. Black, red and green colors refer to the pn, MOS\,1 
and MOS\,2 data sets, respectively. (See the online version of the 
article for the colour figure.)
}
\label{fig:spectrum}
\end{center}
\end{figure}

A blackbody model is rejected by the data ($\chi^2_\nu = 5.2$ 
for 368 degrees of freedom; dof hereafter). The bremsstrahlung 
model yields $\chi^2_\nu =1.54$ for 368 dof, with the temperature 
pegged to the highest allowed value of 200 keV. In the disk model,
the temperature is allowed to have a radial dependence, 
$T(r) \propto r^{-p}$. Fixing $p$ to 0.75 (i.e., reproducing the standard 
geometrically thin and optically thick accretion disk of Shakura \& 
Sunyaev 1973) gives $\chi^2_\nu = 1.30$ for 368 dof. If we leave this 
parameter free to vary we obtain $\chi^2_\nu = 1.18$ for 367 dof, with 
$p=0.68^{+0.01}_{-0.02}$. An $F$-test gives a chance probability of 
$\sim1.1 \times 10^{-9}$, corresponding to a 6.3$\sigma$ improvement.
However, the inferred value for the temperature at the inner radius of 
the disk is $\gtrsim 5.9$ keV, which is quite implausible. We find instead 
$\chi^2_\nu=1.19$ for 368 dof for the power law model. However, 
structured residuals are clearly visible at energy $\lesssim0.6$~keV 
and around 6.6~keV. The inclusion of a partial covering absorber 
(\textsc{pcfabs}), accounting for additional partial absorption, and a 
Gaussian feature (\textsc{gauss}) in emission, leads to an improvement 
in the shape of these residuals and to a more satisfactory modeling of 
the data (see Fig.~\ref{fig:spectrum}). We obtain $\chi^2_\nu=0.97$ for 
363 dof. The best-fitting parameters are listed in Table~\ref{tab:spectra}. 
The interstellar absorption column density, $\nh_{, \rm{ISM}}\sim2\times 
10^{21}$~cm$^{-2}$, is about one order of magnitude lower than the 
total Galactic value in the direction of the source ($\sim1.25\times
10^{22}$~cm$^{-2}$; Willingale et al. 2013), implying a closeby location 
of this source within our Galaxy. The nearby source CTB~87, believed 
to be at a distance of about 6.1 kpc, has a much higher $\nh$ value of 
$1.4\times10^{22}$~cm$^{-2}$ (Matheson et al. 2013), further supporting 
the closer distance of this source. The partial ($\approx 83$\%) column 
density is larger than the interstellar column density by a factor of $\sim 3$ 
(see Table~\ref{tab:spectra}), suggesting that the majority of the contribution 
to the observed absorption is due to an absorber localized close to the source. 
The broad emission feature around 6.6~keV is suggestive of different 
contributions, in particular thermal and fluorescent iron lines. The 0.3--10~keV 
unabsorbed flux is ($3.16\pm0.05$) $\times 10^{-12}$ \flux.

\begin{table}
\begin{minipage}{8.5cm}
\caption{Average spectral fit results for the \xmm\ observation of \cv. pn and MOSs 
spectra were fitted together to the \textsc{tbabs*pcfabs*(pegpwrlw+gauss)} model 
in the 0.3--10~keV energy range. Uncertainties are quoted at the 90\% confidence 
level for a single parameter of interest.}
\label{tab:spectra}
\centering
\begin{tabular}{@{}lc}
\hline
Parameter									& Value \\
\hline \vspace {0.1cm}
$\nh_{, \rm{ISM}}^a$   	($10^{21}$ cm$^{-2}$) 		& $2\pm1$      \\ 	\vspace {0.1cm}
$\nh_{, \rm{pcf}}$   		($10^{21}$ cm$^{-2}$) 		& $6.1_{-0.7}^{+1.0}$      \\ 	\vspace {0.1cm}
Covering fraction		(\%)						& $83_{-16}^{+10}$		\\ 	\vspace {0.1cm}
$\Gamma$									& $1.15\pm0.04$		\\	\vspace {0.1cm}
Energy of line			(keV)					& $6.61\pm0.09$	  	 \\	\vspace {0.1cm}
Width of line		    	(keV)					& $0.24_{-0.08}^{+0.11}$ 	   \\	\vspace {0.1cm}
Normalization of line	   	($10^{-6}$ cm$^{-2}$ s$^{-1}$)	& $5.2_{-1.6}^{+1.9}$	   \\	\vspace {0.1cm}
Equivalent width   		(keV)                  				& $0.18\pm0.06$       	   \\	\vspace {0.1cm}
Absorbed flux$^b$   	     	($10^{-12}$ \flux)			& $2.75_{-0.03}^{+0.05}$     \\	\vspace {0.1cm}
Unabsorbed flux$^b$  	($10^{-12}$ \flux)			& $3.16\pm0.05$	    	    \\	\vspace {0.1cm}
$\chi^2_{\nu}$ 	     (dof)        				         		& 0.97 (363) 		    	     \\	\vspace {0.1cm}
Null hypothesis probability						& $6.5 \times 10^{-1}$    	     \\	
\hline 
\end{tabular}  
\begin{list}{}{}
\item[$^{a}$] The abundances are those of Wilms et al. (2000). The photoelectric absorption cross-sections are from Verner et al. (1996).
\item[$^{b}$] In the 0.3--10~keV energy range. 
\end{list}
\end{minipage}
\end{table}

\subsubsection{Phase-resolved spectral analysis}
\label{pps}

\begin{figure*}
\begin{center}
\includegraphics[width=8cm]{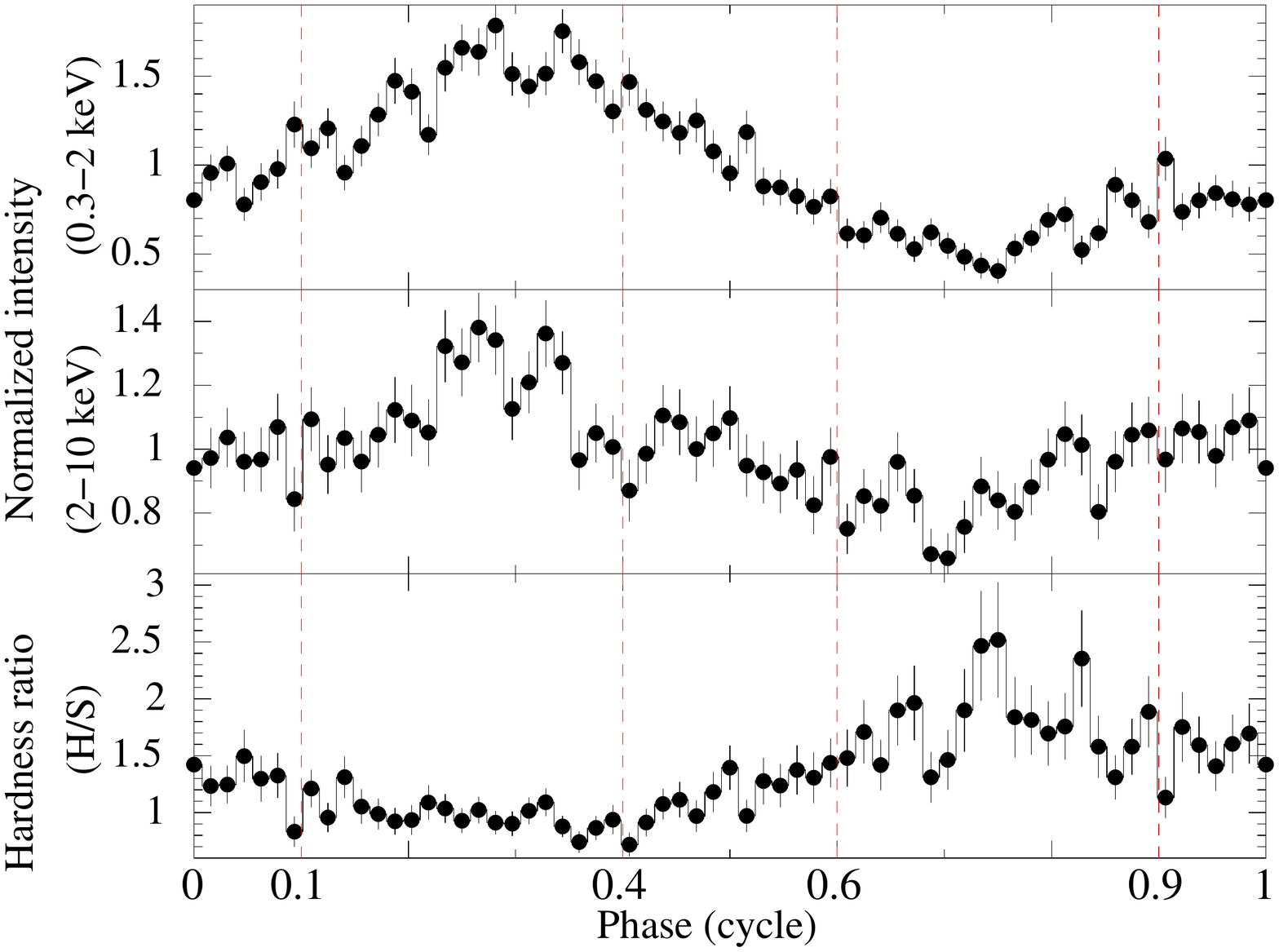}
\includegraphics[width=8cm]{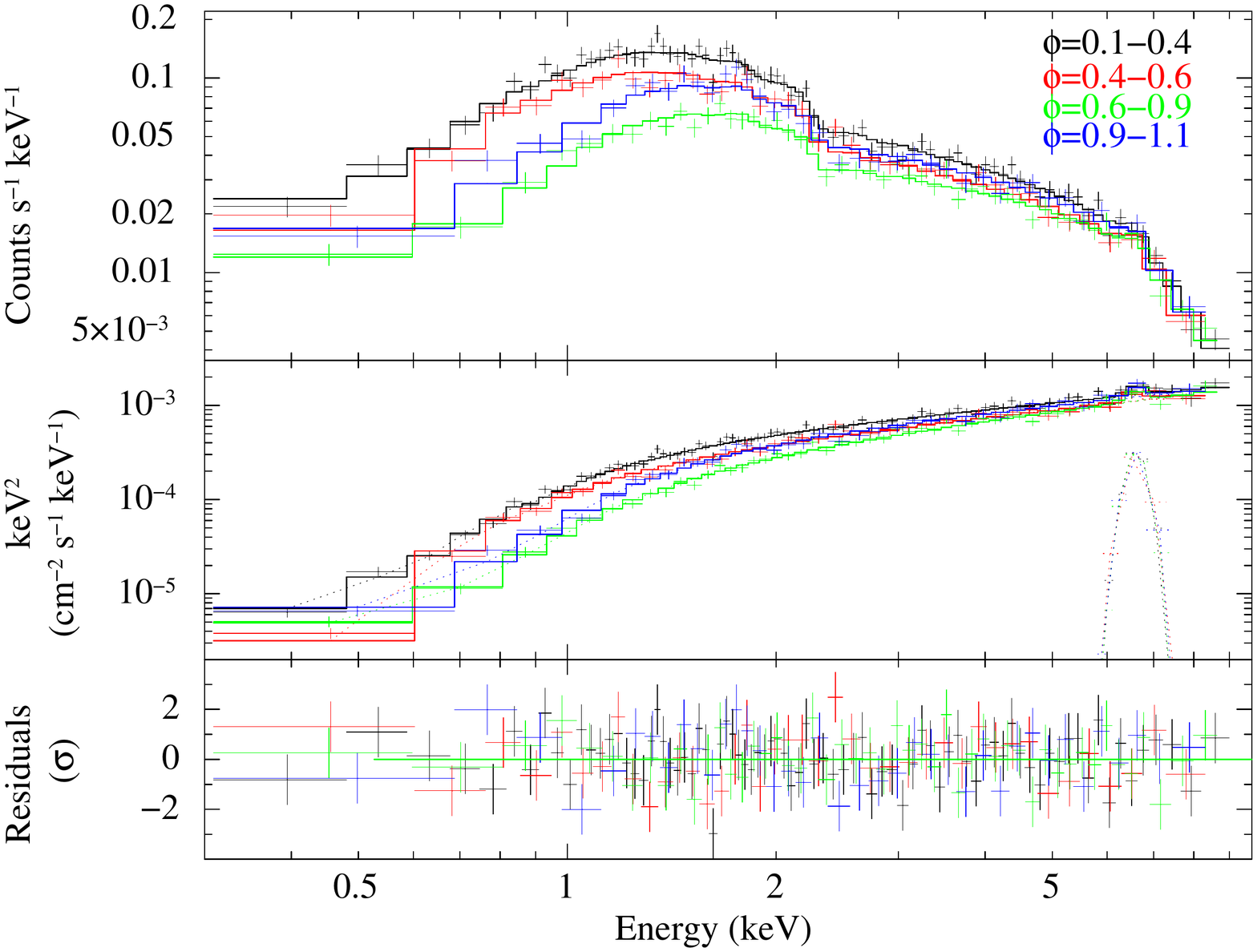}
\vspace{-0.5cm}
\caption{Left-hand panel: 64-bin pulse profile of \cv\ in two different energy 
intervals. The hardness ratio between the hard and soft bands is also plotted 
at the bottom. The red vertical dashed lines indicate the phase ranges used 
for the phase-resolved spectral analysis. Right-hand panel: phase-resolved 
spectra of \cv\ extracted from \xmm\ EPIC-pn data. The solid lines represent 
the best-fitting model to the data (a power law plus a Gaussian feature in emission, 
both corrected for interstellar and local absorption). $E^2\times f(E)$ unfolded 
spectra (middle panel) and post-fit residuals in units of standard deviations 
(bottom panel) are also shown. (See the online version of the article for the colour 
figure).
}
\label{fig:hardness_ratio}
\end{center}
\end{figure*}

The normalized energy versus phase image relative to the 7196-s periodicity 
hints at a phase-variable emission, possibly due to a varying absorption along 
the line of sight (see the left-hand panel of Fig.~\ref{fig:profiles}). To better 
investigate the variability of the X-ray emission along the phase, we computed 
an hardness ratio between the hard (2--10~keV) and soft (0.3--2~keV) counts 
along the cycle (see Fig.~\ref{fig:hardness_ratio}). We then carried out a 
phase-resolved spectroscopy accordingly, selecting the following four phase 
intervals: 0.1--0.4 (corresponding to the softest state, around the maximum of 
the modulation), 0.4--0.6, 0.6--0.9 (related to the hardest state, close to the 
minimum of the modulation) and 0.9--1.1.

We used all EPIC (pn + MOSs) data and fitted the spectra together to the best-fitting 
average model. We fixed the interstellar absorption column density, the power law 
photon index and the centroid and width of the Gaussian feature at the phase-averaged 
values, after having verified that the lower statistics in the phase-resolved spectra did 
not allow us to study possible differences for the values of the parameters of the Gaussian 
feature. We obtained an acceptable $\chi^2_\nu = 1.01$ for 199 dof. Tying up the partial 
column density across the spectra yielded a significantly worse fit ($\chi^2_\nu = 1.50$ 
for 202 dof). Therefore, as shown in Table~\ref{tab:spectra_pps}, the variability of the 
X-ray emission along the phase cycle can be successfully ascribed to changes in both 
the density and spatial extension of the localized absorbing material, and the power law 
normalization. In particular, the partial column density is significantly larger at the minimum 
(about $1.2 \times 10^{22}$ cm$^{-2}$) than at the maximum ($\sim 2.8 \times10^{21}$ 
cm$^{-2}$). Unabsorbed 0.3--10~keV fluxes are ($2.77\pm0.10$) and ($3.62\pm0.09$) 
$\times 10^{-12}$ \flux, respectively (see Table~\ref{tab:spectra_pps}). The phase-resolved 
spectra together with the best-fitting model are shown in the right-hand panel of 
Fig.~\ref{fig:hardness_ratio} (only for pn data for plotting purpose). 

\begin{figure}
\begin{center}
\includegraphics[width=18cm]{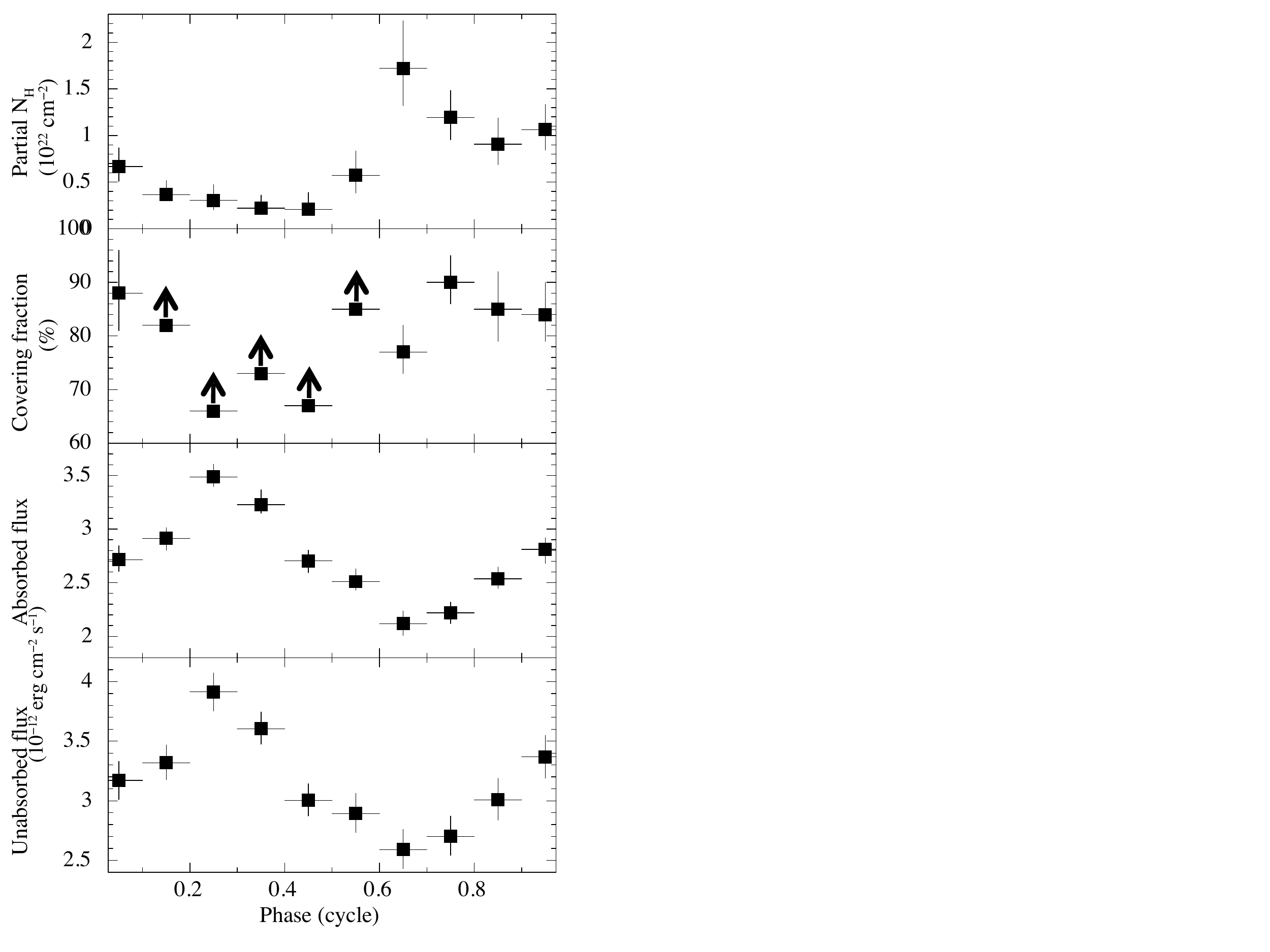}
\caption{Evolution of the column density and covering fraction of the partial absorber 
and of the 0.3--10~keV fluxes as a function of the phase for the \xmm\ observation of \cv. 
The 10 pn spectra (each corresponding to a 0.1-wide phase interval) were fitted together to the 
\textsc{tbabs*pcfabs*(pegpwrlw+gauss}) model in the 0.3--10~keV energy range (see the text 
for details). Uncertainties and lower limits (indicated by arrows) are reported at a 90\% confidence 
level for a single parameter of interest.}
\label{fig:pps_10bins}
\end{center}
\end{figure}

To better characterize the phase-dependence of the parameters, we repeated 
the analysis (only for pn data) by dividing the phase cycle into 10 equal intervals, 
each of width 0.1 in phase, and fitting the corresponding spectra together to 
the same model as above. We obtained $\chi^2_\nu=0.96$ for 404 dof. The 
evolution of the parameters and the fluxes along the cycle is shown in 
Fig.~\ref{fig:pps_10bins}. An anti-correlation between the partial absorption column 
density and the X-ray intensity can be seen in the figure.

\begin{table}
\centering 
\caption{Phase-resolved spectral fit results for the \xmm\ observation of \cv. 
pn and MOSs spectra were fitted together to the \textsc{tbabs*pcfabs*(pegpwrlw+gauss}) 
model in the 0.3--10~keV energy range. The interstellar absorption column density 
and the power law photon index were fixed at the phase-averaged values, resulting 
in $\chi^2_\nu=1.01$ for 199 dof (see the text). Uncertainties and lower limits are 
quoted at a 90\% confidence level for a single parameter of interest.} 
\label{tab:spectra_pps}
\begin{tabular}{@{}lcccc}
\hline
Phase range	  		& $\nh_{, \rm{pcf}}$			& Cvf$^a$ 		& Abs flux$^b$				& Unabs flux$^b$ \\  	
			  		& ($10^{21}$ cm$^{-2}$)		& (\%)			&  \multicolumn{2}{c}{($10^{-12}$ \flux)}		\\
\hline  \vspace {0.1cm}
$0.1-0.4$ (max)  		& $2.8_{-0.4}^{+0.8}$		& $\gtrsim81$		& $3.23_{-0.06}^{+0.07}$		& $3.62\pm0.09$		\\ \vspace {0.1cm}
$0.4-0.6$ 		  		& $3.2_{-0.6}^{+1.3}$		& $\gtrsim77$		& $2.62_{-0.08}^{+0.09}$		& $2.95_{-0.10}^{+0.11}$	 	\\ \vspace {0.1cm}
$0.6-0.9$ (min)	  		& $12.0\pm0.2$			& $84\pm3$		& $2.30_{-0.06}^{+0.07}$		& $2.77\pm0.10$		\\ \vspace {0.1cm}		
$0.9-1.1$ 		  		& $8.2_{-0.1}^{+0.2}$		& $86_{-4}^{+5}$	& $2.78_{-0.09}^{+0.08}$		& $3.27\pm0.12$		 \\ 
\hline						
\end{tabular}
\begin{list}{}{}
\item[$^{b}$] Covering fraction of the partial absorber.
\item[$^{b}$] In the 0.3--10~keV energy range.
\end{list}
\end{table}

\subsection{\axaf}
\label{axaf}

The \axaf\ satellite observed \cv\ twice, on 2001 Jul 8 and 2010 
Jan 16--17 (see Table~\ref{tab:obslog}). The first observation 
(obs. ID 1037; PI: Garmire) was performed with the Advanced 
CCD Imaging Spectrometer spectroscopic CCD array (ACIS-S; 
Garmire et al. 2003) set in faint timed-exposure (TE) imaging 
mode. The total exposure was about 17.8 ks and the source 
was positioned on the back-illuminated S3 chip. The second 
observation (obs. ID 11092; PI: Safi-Harb) used instead the 
imaging CCD array (ACIS-~I) operated in very faint TE mode 
and lasted about 69.3 ks. The source was positioned on the I3 chip.

We analyzed the data following the standard analysis threads\footnote{See http://cxc.harvard.edu/ciao/threads/pointlike.} 
with the \axaf\ Interactive Analysis of Observations software (\textsc{ciao}, 
v. 4.7; Fruscione et al. 2006) and the calibration files in the \axaf\ 
\textsc{caldb} (v. 4.6.9). 

The source fell close to the edge of the CCD in the 2001 observation 
and was very far off-axis in that carried out in 2010 (for both these 
observations the aim point of the ACIS was indeed the supernova 
remnant CTB~87, which is located at an angular distance of $\sim5.2$ 
arcmin from the nominal position of \cv). Source photons were then 
collected from a circular region around the source position with a 
radius of 15 arcsec. Background was extracted from a nearby circle 
of the same size. We converted the photon arrival times to Solar System 
barycentre reference frame using the \textsc{ciao} tool \textsc{axbary} 
and restricted our analysis to photons having energies between 0.3 
and 8~keV. The average source net count rates in this band are 
$(8.5 \pm 0.2) \times 10^{-2}$ and $(1.23 \pm 0.01) \times 10^{-1}$ 
counts~s$^{-1}$ for the first and second observation, respectively. 

We created the source and background spectra, the associated 
redistribution matrices and ancillary response files using the \textsc{specextract} 
script.\footnote{Ancillary response files are automatically corrected to 
account for continuous degradation in the ACIS CCD quantum efficiency.} 
We grouped the background-subtracted spectra to have at least 30 and 
100 counts in each spectral bin for the first and second observation, respectively.

\subsubsection{Timing analysis}

Very recently, timing analysis of the longest \axaf\ observation (obs. ID 11092) unveiled the 
presence of the 2-hr periodicity observed in the \xmm\, data (Halpern \& Thorstensen 
2015). The energy-dependent light curves folded on this period closely resemble the profiles 
derived from the \xmm\ observation (see e.g. Fig. 17 of Halpern \& Thorstensen 2015), with 
a modulation amplitude that  decreases as the energy increases. We confirmed this detection 
in the soft X-ray band (0.5--2~keV) and verified that, except for the harmonic, the other 
periodicities discovered in the \xmm\ observation are not visible, likely due to the lower counting 
statistics in the \axaf\ data sets.

Our searches for periodicities in the data of the other observation (obs. ID 1037) 
were instead inconclusive. Only one power peak is visible and we note that it is 
coincident with the frequency of known artificial signal due to the source dithering 
off the chip at regular time intervals.\footnote{See http://cxc.harvard.edu/ciao/why/dither.html.}

\subsubsection{Spectral analysis}

The values for the source net count rate translate into a pile-up fraction of 
about 10 and 15\% for the first and the second observation, respectively, 
as estimated with \textsc{pimms} (v 4.8).\footnote{http://heasarc.gsfc.nasa.gov/cgi-bin/Tools/w3pimms/w3pimms.pl.}
We then accounted for possible spectral distortions using the model of Davis (2001), 
as implemented in \textsc{xspec}, and following the recommendations in 
`{\em The Chandra ABC Guide to Pile-up}'.\footnote{See http://cxc.harvard.edu/ciao/download/doc/pile-up$_-$abc.pdf.}
We fitted the two spectra together in the 0.3-8~keV energy range with an absorbed 
power law model and all the parameters left free to vary. However, the column density 
was compatible between the two epochs at the 90\% confidence level and 
thus was tied up. We obtained $\chi^2_\nu=1.21$ for 118 dof, with \nh = $(6\pm1) 
\times 10^{21}$~cm$^{-2}$. The spectrum was harder in 2001 ($\Gamma = 0.6 \pm 
0.1$) than in 2010 ($\Gamma = 0.95 \pm 0.06$). The emission feature around 6.6~keV 
is not detected, likely due to the lower statistics compared to the \xmm\ data sets. 
The addition of a Gaussian feature in emission with parameters fixed at the values 
derived from the \xmm\ observation (see Table~\ref{tab:spectra}) yields upper limits for 
the equivalent width of 52 and 119 eV in 2001 and 2010, respectively (at a 90\% 
confidence level). The 0.3--10~keV unabsorbed fluxes at the two epochs were $(9.0\pm0.6) 
\times 10^{-12}$ and $(4.05\pm0.09) \times 10^{-12}$ \flux. The soft X-ray flux of \cv\ was 
thus larger in 2001 than in 2010 and 2014 (the epoch of the \xmm\ observation) by a factor 
of $\sim2.2$ and $\sim2.8$, respectively.

\subsection{\swift}
\label{swift}

Our analysis of \swift\ data is mainly aimed at comparing the simultaneous soft X-ray 
and ultraviolet/optical fluxes of \cv\ at different epochs. Therefore, we focus here on the 
two longest observations of the source (obs IDs: 00035639003, 00041471002; see 
Table~\ref{tab:obslog}). The X-ray Telescope (XRT; Burrows et al. 2005) was set in photon 
counting mode in both cases. 

We processed the data with standard screening criteria and generated 
exposure maps with the task \textsc{xrtpipeline} (v. 0.13.1) from the 
\textsc{ftools} package. We selected events with grades 0--12 and extracted 
the source and background event files using \textsc{xselect} (v. 2.4). We 
accumulated the source counts from a circular region centered at the peak 
of the source point-spread function and with a radius of 20 pixels (one XRT 
pixel corresponds to about 2.36 arcsec). To estimate the background, we 
extracted the events within a circle of the same size sufficiently far from the 
blazar and other point sources. 

We created the observation-specific ancillary response files (using 
exposure maps) with \textsc{xrtmkarf}, thereby correcting for the loss 
of counts due to hot columns and bad pixels and accounting for different 
extraction regions, vignetting and PSF corrections. We then assigned 
the redistribution matrices v012 and v014 available in the \textsc{heasarc} 
calibration database to the 2006 and 2010 data sets, respectively.\footnote{See 
http://www.swift.ac.uk/analysis/xrt/rmfarf.php.} We grouped the spectral channels 
to have at least 20 counts in each spectral bin and fitted in the 0.3 to 10~keV 
energy interval with an absorbed power law model. The inferred 0.3-10~keV 
unabsorbed fluxes are $\sim 6.9$ and $4.1 \times 10^{-12}$ \flux\ in 2006 and 
2010, respectively. This is in line with a fading trend of the source in about 10~yr.

\section{UV/optical observations and data analysis} 
\label{uv}

\cv\ was not positioned on the detectors of the \xmm\ Optical/UV Monitor telescope 
(Mason et al. 2001) throughout the observation. We thus focused on the two 
observations by the Ultra-violet and Optical Telescope (UVOT; Roming et al. 2005) 
aboard \swift, which lasted about 7.0 and 7.2~ks, respectively, and were carried out 
in image mode. All the available filters were used in both cases (see Table~\ref{tab:uvotlog}), 
providing a wavelength coverage within the 1700--6000~\AA\ range. We performed 
the analysis for each filter and on the stacked images. First, we ran the \textsc{uvotdetect} 
task and found the UV and optical counterpart of \cv\ at position $\rm RA=20^h15^m36\fs959$, 
$\rm Dec= +37^\circ11'22\farcs70$ (J2000.0), which is compatible with the location 
reported by Halpern et al. (2001) within the errors. We then used the \textsc{uvotsource} 
command, which calculates detection significances, count rates corrected for 
coincidence losses and dead time of the detector, flux densities and magnitudes 
through aperture photometry within a circular region, and applies specific corrections 
due to the detector characteristics. We adopted an extraction radius of 5 and 3~arcsec 
for the UV and optical filters, respectively, and applied the corresponding aperture 
corrections. The derived values for the absorbed magnitudes at the two different 
epochs are listed in Table~\ref{tab:uvotlog} (magnitudes are expressed in the Vega 
photometric system; see Poole et al. 2008 for more details and Breeveld et al. 2010, 
2011 for the most updated zero-points and count rate-to-flux conversion factors). The 
source is brighter in all bandpasses in 2006 than in 2010, confirming the fading of the 
source also in the optical and UV.

To estimate the  X-ray--to--optical flux ratio at the two different epochs of the \swift\ 
observations, we determined the unabsorbed 2--10~keV fluxes measured by the 
XRT and considered the values for the $V$-band magnitudes observed by the UVOT
(central wavelength of 5468 \AA\ and full-width at half-maximum of 769 \AA). 
Adopting the value of the interstellar hydrogen column density derived from our 
model fit to the phase-averaged X-ray spectrum, and a conversion factor of 
$\nh/A_{\rm V}$ of $(2.87 \pm 0.12) \times 10^{21}$~cm$^{-2}$~mag$^{-1}$ 
(according to the relation of Foight et al. 2015), we estimated an optical 
extinction $A_{\rm V} = 0.60\pm0.12$~mag. We note that this value is lower 
than the integrated line-of-sight optical extinction at the position of the source, 
$A_{\rm V} \sim 5.0$~mag, computed according to the recalibration (Schlafly \& 
Finkbeiner 2011) of the extinction maps from Schlegel, Finkbeiner \& Davis (1998). 
Using the Vega magnitude-to-flux conversion, we inferred dereddened $V$-band 
fluxes for the source of $(4.6 \pm 0.8) \times 10^{-12}$ \flux\ in 2006 and $(2.3 \pm 0.4) 
\times 10^{-12}$ \flux\ in 2010. The X-ray--to--optical flux ratio is then 
$F_{\mathrm{X}}/F_{\mathrm{V}} = 1.4 \pm 0.2$ and $1.7 \pm 0.2$, respectively 
(at the 1$\sigma$ confidence level), thus compatible between the two epochs.

\begin{table}
\centering
\caption{\swift\ UVOT observations of \cv\ at two different epochs. Magnitudes are in the Vega 
photometric system and are not corrected for interstellar extinction. Uncertainties are quoted at 
a 1$\sigma$ confidence level, whereas upper limits are given at a 3$\sigma$ confidence level.}
\label{tab:uvotlog}
\begin{tabular}{@{}cccc}
\hline
Filter								& Date			& Exposure       & Magnitude           \\
 								&				& (s) 			& (mag) \\
\hline
\multirow{2}{*}{$UVW2$ }				& 2006 Nov 17		& 2421	 	 & $19.76\pm0.14$ \\
								& 2010 Aug 6--10	& 944		 & $19.96\pm0.25$ \\
\hline
\multirow{2}{*}{$UVM2$ } 				& 2006 Nov 17 		& 1567	 	 & $19.83\pm0.21$ \\
								& 2010 Aug 6--10	& 4940		 & $21.22\pm0.36$ \\
\hline
\multirow{2}{*}{$UVW1$ }				& 2006 Nov 17		& 1210	 	 & $18.51\pm0.09$ \\
								& 2010 Aug 6--10	& 629		 & $>19.98$ \\
\hline
\multirow{2}{*}{$U$ }					& 2006 Nov 17		& 598	 	 & $17.65\pm0.06$ \\
								& 2010 Aug 6--10	& 298		 & $18.95\pm0.14$ \\
\hline							
\multirow{2}{*}{$B$ }					& 2006 Nov 17		& 598	 	 & $18.00\pm0.05$ \\
								& 2010 Aug 6--10	& 236		 & $19.28\pm0.14$ \\
\hline
\multirow{2}{*}{$V$ }					& 2006 Nov 17		& 598	 	 & $17.19\pm0.06$ \\
								& 2010 Aug 6--10	& 182		 & $17.94\pm0.13$ \\	
\hline
\end{tabular}
\end{table}

\section{Discussion}
\label{discussion}

Based on the wealth of observations we have collected for \cv, from the optical to the 
X-rays, we can now explore the possibility that this source indeed belongs to the class 
of mCVs, as proposed by Halpern \& Thorstensen (2015).

\subsection{The X-ray periodicities and the accretion geometry}

The X-ray properties of mCVs are strongly related to the accretion flow onto the WD. 
The accretion configuration is different for the two subclasses of mCV: in polars the 
high magnetic fields prevent the formation of an accretion disk and material flows in 
a column-like funnel; in IPs matter accretes either via a truncated disc (or ring; Warner 
1995) or via a stream as in the polars (the diskless accretion model first proposed by 
Hameury, King \& Lasota 1986). Whether the disk forms, or not, close to the WD surface, 
the infalling material is magnetically channeled towards the polar caps, resulting in 
emission strongly pulsed at the spin period of the accreting WD in disk accretors or 
at the beat period in disk-less accretors. Matter is de-accelerated as it accretes, and a 
strong standing shock develops in the flow above the WD surface. X-rays are then 
radiated from the post-shock plasma. 

Within the mCV scenario, the $\sim$\,2-hr modulation discovered in the X-ray 
(0.3--10~keV) emission of  \cv\ (see \ref{timing}) can be naturally interpreted as the 
trace of the accreting WD spin period. The spin modulation has a highly structured 
shape that resembles those observed in IPs rather than in polars, which show instead 
a typical on/off behaviour due to self occultation of the accreting pole and thus 
alternate bright and faint phases (see Warner 1995; Matt et al. 2000 for the prototypical 
polar; Ramsay et al. 2009, Traulsen et al. 2010 and Bernardini et al. 2014 for more 
recent studies). The modulation has a large amplitude (about 30\%) and the 
detection of the second harmonic suggests that accretion occurs onto two polar regions 
located at the opposite sides of the WD surface, and both visible. The main accreting 
pole produces the sinusoidal modulation, whereas the other pole is responsible for 
the secondary hump observed around the minimum of the spin pulse profile (see 
Fig.~\ref{fig:profiles}). The relative contribution of the two poles to the observed 
emission can be quantitatively estimated from the fundamental-to-second harmonic 
amplitude ratio, which is $\sim 3$ for this system. 

The additional weaker periodicities detected in the power spectrum can be identified 
as the orbital sidebands of the spin pulse in an IP system. Theory predicts that, if accretion 
onto the WD occurs via a funnel rather than a disk, the asynchronous rotation of the 
WD should produce modulation not only at the spin frequency ($\omega$), but also 
at the orbital ($\Omega$) and beat ($\omega$ -- $\Omega$) frequencies. Higher 
harmonics of the beat frequency may be present for specific combinations of the system 
inclination and the offset angle between the magnetic and spin axes (see Wynn \& 
King 1992). Setting the spin frequency as $\omega= 1.39  \times 10^{-4}$~Hz, the 
frequencies $9.4 \times 10^{-5}$, $1.20 \times 10^{-4}$, $1.65  \times 10^{-4}$ and 
$2.6 \times 10^{-4}$~Hz (at which excess of power is observed, see \ref{timing}) would 
correspond to the sideband frequencies $\omega$ -- 2$\Omega$, $\omega$ -- $\Omega$, 
$\omega$ + $\Omega$ and 2$\omega$ -- $\Omega$, respectively, for a putative 
binary orbital frequency $\Omega \sim 2.26 \times 10^{-5}$~Hz (i.e. an orbital period 
$P_{{\rm orb}} \sim 12$~hr). We note that the X-ray power spectrum does not show any 
evident signal at this putative period, which is not surprising given the low amplitude 
of the sideband $\omega$ -- $\Omega$ and $\omega$ + $\Omega$ signals ($\sim 6$ 
and 8\%, respectively).

Although the prominent peak at the spin frequency in the X-ray power spectrum implies 
that accretion occurs predominantly through a Keplerian disk (the circulating material 
loses all knowledge of the orbital motion; see Wynn \& King 1992), the detection of the 
orbital sidebands suggests that a non negligible fraction of matter leaps over the disk 
and is directly channeled onto the WD magnetic poles within an accretion column. This 
sort of hybrid accretion mode, dubbed `disk-overflow', is supported by theoretical 
simulations (see e.g. Armitage \& Livio 1998) and it has been observed in other confirmed 
IPs (see Bernardini et al. 2012, and references therein). 
The spin-to-beat amplitude ratio of $4.8\pm0.4$ (at the 90\% confidence level) implies 
that approximately 80\% of matter is accreted through the disk-fed mode. 

The spin modulation amplitude decreases as the energy increases (the pulsed fraction is 
only about 9\% above 5~keV). This behaviour is typical for IPs. In these systems 
the softest X-rays are more prone to absorption from neutral material in the pre-shock flow, 
whereas X-rays at higher energies are essentially unaffected by photo-electric absorption, 
resulting in a decrease of the modulation amplitude (Rosen, Mason \& Cordova 1988).
Furthermore, the overall pulse shape changes with time (see the right-hand panel of 
Fig.~\ref{fig:profiles}), in particular the intensity of the secondary peak increases. Such 
variations are also observed in a few similar systems (e.g. Bernardini et al. 2012; Hellier 2014),
and are likely due to the different accretion modes occurring in these binaries.

\subsection{The X-ray spectral properties}

The average X-ray spectrum of \cv\ can be described by an absorbed hard power law ($\Gamma= 
1.15\pm0.04$) plus a broad ($\sim0.25$~keV) Gaussian emission feature at $6.61\pm0.09$~keV. 
The absorption is provided by the interstellar medium component (with column density $\nh_{, \rm{ISM}} 
\sim 2 \times 10^{21}$~cm$^{-2}$) plus a thicker ($\nh_{, \rm{pcf}} \sim 6 \times 10^{22}$~cm$^{-2}$) 
absorber partially covering the X-ray emitting regions. A partial covering absorption and the iron 
complex are nevertheless defining properties of mCVs (e.g., Yuasa et al. 2010; Girish \& Singh 2012; 
Ezuka \& Ishida 1999). The large width for the line is suggestive of both thermal and fluorescent 
contributions (i.e. the 6.67~keV He-like and the 6.97~keV H-like lines, plus the 6.4~keV fluorescent 
line). In particular the fluorescent line is produced by reflection of X-rays from cold matter that could 
be the WD surface or the pre-shock flow. 

We observe a conspicuous hardening at the minimum of the rotational phase cycle and we find 
that the phase-variable emission can be accounted for mainly by changes in the amount of local 
absorbing material and the power law normalization. Moreover, the partial column density is 
anti-correlated with the X-ray flux along the cycle. The variability in the extension of the local 
absorber (covering fraction) is instead less significant and only lower limits can be inferred around 
the maximum of the modulation. Although localised absorption is also detected in the X-ray light 
curves of polars, this is generally pronounced during the bright phase where the typical dips are due 
to absorption when the accretion column points towards the observer.

The spectral changes along the modulation phase in \cv\ are typically observed in IPs and can be 
explained within the accretion curtain scenario presented by Rosen et al. (1988). According to their 
model, the modulation is mainly caused by spin-dependent photoelectric absorption from pre-shock 
material flowing from the disk towards the polar caps in arc-shaped structures extending above the 
WD. In this context, the spin-phase modulation of \cv\ in the soft X-rays likely reflects changes in the 
absorption along the curtains.
At the minimum of the rotational phase the softest X-ray photons from the WD surface are highly 
absorbed, resulting in a spectral hardening and a decrease in the amount of emission. On the other 
hand, at the maximum of the modulation, the absorption is less efficient and the radiation intensity will 
be larger.

\subsection{The long-term X-ray variability and multiwavelength emission}

\cv\ appears to have gradually faded in the X-rays since 2001 (see \ref{axaf} and \ref{swift}), 
a behaviour that is not uncommon in mCVs of the IP type (Warner 1995).

The simultaneous X-ray and UV/optical observations of \cv\ carried out with \swift\ enable 
a straight estimate of the flux in different energy bands (see Section~\ref{uv}). The source 
appears to be brighter in 2006 November than in 2010 August in all bands. The ratio between 
the unabsorbed X-ray and optical flux is roughly the same at the two epochs: 
$F_{\mathrm{X}}/F_{\mathrm{V}} \sim 1.4-1.7$. If we interpret the X-ray luminosity as accretion 
luminosity, the rather stable ratio suggests that both the X-ray and optical luminosities are 
tracing changes in the mass accretion rate. Moreover, a relatively low value for the ratio 
is not surprising for the CVs, which have X-ray--to--optical flux ratios much lower with 
respect to LMXBs (see van Paradijs \& Verbunt 1984; Patterson \& Raymond 1985; Motch et 
al. 1996), with the mCVs showing higher ratios compared to non-magnetic systems (see 
Verbunt et al. 1997). Therefore, both the intrinsic emission of the disk and the X-ray reprocessed 
radiation provide an important contribution to the UV and optical emission in these systems.

\subsection{The system parameters}

The 2-hr spin pulsations pin down \cv\ as the second slowest rotating WD in the class of IPs
(after RX\,J0524+4244, also known as "Paloma", with $P_{\rm spin} \sim 2.3$~hr; see Schwarz et al. 2007).
If the additional peaks detected in the X-ray power spectrum are indeed the orbital sidebands, 
then the estimated orbital period of the system of about 12~hr would imply a degree of asynchronism 
$P_{{\rm spin}} / P_{\rm{orb}} \sim 0.16$, which is larger than that observed for most of the long-orbital 
period IPs (Norton et al. 2004; Bernardini et al. 2012; see also the catalogue available at the Intermediate 
Polar Home Page\footnote{http://asd.gsfc.nasa.gov/Koji.Mukai/iphome/catalog/members.html}). 
Taken at face value, the long spin period would be peculiar when compared to other slow rotators 
which are all short orbital period systems, including the IP Paloma.

For a system with orbital period of about 12~hr, the likely companion would be a G- or early 
K-type star (Smith \& Dhillon 1998). Adopting an absolute magnitude in the $K$-band in the 
range 3.50--4.51~mag (Bilir et al. 2008), using the value for the $K$-band magnitude of \cv\ 
($14.79 \pm 0.15$~mag), and assuming that the donor star in this binary system totally contributes 
at these wavelengths, the estimated distance would be in the range 1--2~kpc, with no intervening 
absorption. Our estimate does not change significantly if the effects of interstellar absorption are 
considered, since adopting the optical reddening derived in Section~\ref{uv} and the extinction 
coefficients at different wavelengths of Fitzpatrick (1999), the inferred absorption in this band is 
only $A_{\rm K} \sim 0.07$ mag.

\section{Conclusions}
\label{conclusion}

A deep \xmm\ observation of \cv, together with archival \axaf\ and \swift\ observations, allowed 
us to unambiguously identify this source with a magnetic cataclysmic variable. 

Although we did not detect directly the orbital period of the system, several properties point toward the 
intermediate polar identification, and the 2\,hr modulation as the white dwarf spin period. In particular, no 
coherent signal is observed at shorter periods (ruling out the presence of a lower spin period\footnote{Note 
that to date no IP has been observed with a detected orbital period, but no evidence for the spin period.}), 
and the presence of additional weaker periodicities close to the most significant one at 2\,hr could be reconciled 
with orbital sidebands for a putative orbital period of about 12~hr. In this case the system would be an 
intermediate polar with an anomalous spin--to--orbit period ratio for its long orbital period. Alternatively, the 
orbital period could be one of the sideband periods itself, and in this case \cv\, would resemble systems like 
Paloma (Schwarz et al. 2007) or IGR\,J1955$+$0044 (Bernardini et al. 2013), which are characterized by a 
high spin--to--orbit period ratio ($P_{{\rm spin}} / P_{\rm{orb}} \sim 0.9$) and are believed to be intermediate 
polars (or pre-polars) on their way to reach synchronism. 

Both the shape and the spectral characteristics of the 2\,hr modulation are typical of the intermediate polars. 
Furthermore, the amplitudes of the different modulations indicate that accretion takes place onto the two polar caps 
mostly through a truncated disk (about 80\%), with a non negligible contribution from an accretion stream, again 
a peculiarity shared by several intermediate polars.

Further optical observations are planned to shed light on the exact orbital period of this intriguing system, 
which in any case appears to play a key role in understanding of the evolution of magnetic cataclysmic variables.

\section*{Acknowledgements} 
The scientific results reported in this article are based on observations obtained with \xmm\ and \swift\
and on data obtained from the \axaf\ Data Archive. \xmm\ is an ESA science mission with instruments 
and contributions directly funded by ESA Member States and NASA. \swift\ is a NASA mission with 
participation of the Italian Space Agency and the UK Space Agency. This research has made use of 
software provided by the \axaf\ X-ray Center (CXC, operated for NASA by SAO) in the application package 
CIAO, and of softwares and tools provided by the High Energy Astrophysics Science Archive Research 
Center (HEASARC), which is a service of the Astrophysics Science Division at NASA/GSFC and the High 
Energy Astrophysics Division of the Smithsonian Astrophysical Observatory. The research has also made 
use of data from the Two Micron All Sky Survey, which is a joint project of the University of Massachusetts 
and the Infrared Processing and Analysis Center/California Institute of Technology, funded by NASA and 
the National Science Foundation. FCZ and NR are supported by an NWO Vidi Grant (PI: Rea) and by the 
European COST Action MP1304 (NewCOMPSTAR). NR, AP and DFT are also supported by grants 
AYA2012-39303 and SGR2014-1073. DdM acknowledges support from INAF--ASI I/037/12/0. AP is supported 
by a Juan de la Cierva fellowship. SSH acknowledges support from NSERC through the Canada Research 
Chairs and Discovery Grants programs, and from the Canadian Space Agency. FCZ thanks M.~C. Baglio and 
P.~D'Avanzo for useful discussions. We thank the anonymous referee for comments on the manuscript.

\bsp

\label{lastpage}
\end{document}